\documentclass[12pt]{article}
\usepackage[left=2cm,top=2cm,right=2cm,nohead,foot=1cm]{geometry}

\usepackage{subfigure}
\usepackage{graphicx}
\usepackage{amssymb}
\usepackage{epstopdf}
\usepackage[left=2cm,top=2cm,right=2cm,nohead,foot=1cm]{geometry}

\newcommand{\ignore}[1]{}
\newcommand{\be}{\begin{equation}} \newcommand{\ee}{\end{equation}}
\newcommand{\ba}{\begin{eqnarray}} \newcommand{\ea}{\end{eqnarray}}
 \renewcommand{\bf}{\textbf}

\renewcommand{\a}{\alpha} \renewcommand{\b}{\beta}

\newcommand{\p}{\partial}

\begin{document}

%%%%%%%%%%%%%%%%%%%%%%%%%%%%%%%%%%%%%%%%%%%%%%%%%%%%%%%%%%%%%%%%%

\input{epsf}

\title{Signatures of Pseudoscalar Photon Mixing in CMB Radiation}

\author{Nishant Agarwal$^a$, Pankaj Jain$^b$, Douglas W. McKay$^c$ and John P. Ralston$^c$
\\
$^a$Department of Astronomy, Cornell University, Ithaca, NY - 14853, USA\\
$^b$Physics Department, IIT, Kanpur - 208016, India\\
$^c$Department of Physics \& Astronomy, University of Kansas,\\ Lawrence, KS - 66045, USA\\
nagarwal@astro.cornell.edu;
pkjain@iitk.ac.in;
dmckay@ku.edu;
ralston@ku.edu
}

\maketitle

\noindent
\bf {Abstract:}
We model the effect of  photon and ultra-light pseudoscalar mixing on the propagation of electromagnetic radiation through the extragalactic medium. The medium is modelled as a large number of magnetic domains, uncorrelated with one another. We obtain an analytic expression for the different Stokes parameters in the limit of small mixing angle. The different Stokes parameters are found to increase linearly with the number of domains. We also verify this result by direct numerical simulations. We use this formalism to estimate the effect of pseudoscalar-photon mixing on the Cosmic Microwave Background (CMB) polarization. We impose limits on the model parameters by the CMB observations. We find that the currently allowed parameter range admits a CMB circular polarization up to order $10^{-7}$.

%%%%%%%%%%%%%%%%%%%%%%%%%%%%%%%%%%%%%%%%%%%%%%%%%%%%%%%%%%%%%%%%%

%%%%%%%%%%%%%%%%%%%%%%%
\section{Introduction}
%%%%%%%%%%%%%%%%%%%%%%%

Pseudoscalars with very small mass and very weak coupling with visible matter arise in many extensions of the standard model. The most well known example is the axion \cite{PQ,Weinberg, McKay, kim1,Dine,McKay2,Kim}. The subject of pseudoscalar-photon mixing in background magnetic field has been studied by many authors \cite{Karl,PS83,Maiani,RS88,Bradley,CG94,Ganguly}. The mixing changes the intensity as well as the state of polarization of the photons. This phenomenon has also been used to search for very low mass pseudoscalars by laboratory experiments as well as by astrophysical observations. This search has put stringent limits on the mass and coupling parameters of such pseudoscalars \cite{PDG06,Rosenberg,Brockway,astro,CAST,jaeckel,zavattini,Raffelt}. The astrophysical and cosmological consequences have been studied extensively in the literature \cite{Raffelt,Mohanty,Csaki,PJ,sroy,MirizziCMB,MirizziTEV,Song,Gnedin}. 

The polarization of electromagnetic waves can also change due to the presence of a background pseudoscalar field. This phenomenon also has many astrophysical consequences \cite{HS92,Das}.

The propagation through intergalactic medium has been studied earlier by several authors \cite{Csaki,sroy,MirizziCMB,MirizziTEV}. The medium is expected to be turbulent and it is reasonable to model it as a large number of uncorrelated magnetic domains. The magnetic field in a particular domain is assumed to be uniform but points in different, random, directions as we move from one domain to another. Earlier studies were primarily interested in the effect of mixing on the intensity of the wave. In the present paper we are interested in determining the effect on all of the Stokes parameters. This has a wide range of astrophysical and cosmological applications in radio \cite{PJ,anisotropy}, CMB \cite{Raffelt,MirizziCMB,Liu} and optical \cite{PJ,huts,Payez,Piotrovich}. Polarization is a particularly sensitive probe of the pseudoscalar-photon mixing \cite{PJ,sudeep}. We assume that all the magnetic domains have equal length and the plasma density is assumed to be uniform throughout. A detailed analysis of the mixing,
taking into account turbulent plasma density, is given in \cite{CG94,PJ}.

We apply our results to determine the effect on cosmic microwave background radiation. The mixing affects the intensity of radiation as well as the linear and circular polarization. The mixing also results in a distortion of the cosmic microwave background (CMB) spectrum. CMB observations \cite{all} can lead to stringent limits on the pseudoscalar-photon mixing, given current estimates of the intra-galactic magnetic fields and plasma density.

We first set up the basic equations \cite{sudeep}. We consider electromagnetic waves propagating in a background magnetic field. The action can be written as,
\ba 
	S = \int d^{4}x \bigg[ - \frac{1}{4} F_{\mu \nu}F^{\mu \nu} + {g_{\phi}\over 4}  \phi \, \epsilon_{\mu \nu \a \b} F^{\mu\nu}F^{\a \b}  \nonumber \\ 
	+ j_{\mu}A^{\mu} + \frac{1}{2}\p_{\mu}\phi \p^{\mu}\phi - V(\phi)\bigg]. 
	\label{Ldefined2}  
\ea
Maxwell's equations excluding the effects due to gravitation are given by:
\begin{eqnarray}
	\nabla\cdot {\vec E} & = & g_{\phi}\nabla\phi\cdot{(\vec {\cal B}+\vec B )} +\rho ;
	\label{emeq1} \\
	\nabla\times\vec E + {\partial {\vec B }\over \partial t} &=& 0 ;
	\label{emeq2} \\
 	\nabla\times\vec B  - {\partial \vec E \over \partial t} & = & g_{\phi} \left( \vec E \times \nabla \phi - {(\vec {\cal B}+ \vec B )} {\partial \phi\over \partial t} \right)  + \vec j ; 
 	\label{emeq3} \\
	\nabla\cdot{ \vec B } & = & 0.
	\label{emeq4}
\end{eqnarray}
Here $B_{i}$ and $E_{i}$ are the usual magnetic and electric fields and $\vec{\cal B}$ is the background magnetic field. We assume that $\vec{\cal B}$ is independent of space and time and hence set its derivatives to zero.

The pseudoscalar field's equation of motion is
\begin{equation}
	{\partial^2 \phi\over \partial t^2} - \nabla^2\phi + m_\phi^2\phi = - g_{\phi}{\vec E} \cdot (\vec {\cal B} + \vec B).
\label{pseudoscalar}
\end{equation}
We choose the coordinate system such that the $z$-axis points along the direction of propagation and the $x$-axis is parallel to the transverse component of the background magnetic field ${\cal \vec B}$. As shown in Ref. \cite{sudeep}, the longitudinal component of the background magnetic field plays a negligible role and can be ignored. We assume a quasi-monochromatic wave and hence the terms which are quadratic in the fluctuating fields give negligible contribution.
We may write the field equations as
\begin{equation}
	(\omega^2 + \partial_z^2)\left(\matrix{A_\parallel(z)\cr \phi(z)}\right) 
	- M \left(\matrix{A_\parallel(z) \cr \phi(z)} \right) = 0,
	\label{eq:mixing}
\end{equation}
where $\vec A =\vec E/\omega$ and $A_\parallel$ refers to the component parallel to the transverse background magnetic field. The ``mass matrix'' or ``mixing matrix'' is
\begin{equation}
	M = \left(\matrix{\omega_P^2 & - g_{\phi}{\cal B}_T\omega\cr  
	- g_{\phi}{\cal B}_T\omega & m_\phi^2}\right)\ ,
	\label{eq:massmatrix}
\end{equation}
where $\omega_P$ is the plasma frequency, $m_\phi$ the pseudoscalar mass and ${\cal B}_T$ is the magnitude of the transverse component of $\vec {\cal B}$. The plasma frequency is given in terms of the electron density, $n_{e}$, by
\begin{equation}
	\omega_{P}^{2} = \frac{4 \pi \alpha n_{e}}{m_{e}}.
\end{equation}

The component of $\vec A$ along the $y$-axis does not mix with the pseudoscalar field. The matrix $M$ is diagonalized by an orthogonal transformation, $OMO^T = M_D$, where $M_D$ is diagonal. We parameterize the orthogonal matrix,
\begin{eqnarray*}
	O = \left(\matrix{\cos\theta & -\sin\theta\cr \sin\theta & \cos\theta}\ \right).
\end{eqnarray*}
The angle $\theta$ is given by
\begin{equation}
	\tan 2\theta   =  l g_{\phi}{\cal B}_T,
\end{equation}
where the symbol $l$ denotes the oscillation length,
\begin{equation}
	l = \frac{2\omega}{\omega_{P}^{2} - m_{\phi}^{2}}.
\label{eq:oscillationlen}
\end{equation}
We assume that the pseudoscalar mass is very small, $m_\phi \ll \omega_P$. If the mass is much heavier than this, then their mixing with photons produces negligible effect for intergalactic propagation for the range of allowed parameters. The two eigenvalues of the matrix $M$, $\mu_\pm$, may be expressed as,
\begin{equation}
	\mu_{\pm}^{2} = \frac{\omega_{P}^{2} + m_{\phi}^{2}}{2} 
	\pm \frac{1}{2} \sqrt{(\omega_{P}^{2} - m_{\phi}^{2})^2 + (2 g_{\phi}{\cal B}_T \omega)^{2}}.
\end{equation}

%%%%%%%%%%%%%%%%%%%%%%%%%%%%%%%%%%%%%%%%%%%%%%%%%%%%%%%%%%%%%%%%%

%%%%%%%%%%%%%%%%%%%%%%%%%%%%%%%%%%%%%%%%%%%%%%%%%%%%%%
\section{Basic equations of polarization propagation}
%%%%%%%%%%%%%%%%%%%%%%%%%%%%%%%%%%%%%%%%%%%%%%%%%%%%%%

The basic field equations, given in the Introduction, are solved by the procedure described in Ref. \cite{sudeep}. Here we simply quote the final results for propagation of the correlation functions, which are the density matrix elements of the polarization. The most general form of the different correlation functions after propagation through distance $z$, written out explicitly as coupled equations in Appendix A, can be summarized in matrix form as
\begin{equation}
	\rho(z) = P(z) \rho(0) P(z)^{-1},
\label{eq:rhoz}
\end{equation}
where the matrix $\rho(0)$ is defined by

\begin{equation}
	\rho(0) = \left(\matrix{<A_{||}(0) A_{||}^{*}(0)> & <A_{||}(0) A^{*}_{\bot}(0)> & <A_{||}(0) \phi^{*}(0)>\cr <A_{\bot}(0) A^{*}_{||}(0)> & <A_{\bot}(0) A^{*}_{\bot}(0)> &  < A_{\bot}(0) \phi^{*}(0)>\cr <\phi(0) A^{*}_{||}(0)> & < \phi(0) A^{*}_{\bot}(0)> & <\phi(0) \phi^{*}(0)>} \right)\ ,
\label{eq:rho0}
\end{equation}
and the unitary matrix $P(z)$, the solution to the field equations for a given mode $\omega$, is given in the case that the coordinates are chosen so that the ``parallel" axis lies along the transverse component of the external field, ${\vec{\cal B}}_T$, by

\begin{equation}
	P(z) = e^{i(\omega+\Delta_{A})z} \left(\matrix{1-\gamma \sin^{2}\theta & 0 &
\gamma \cos\theta \sin\theta\cr 0&e^{-i(\omega+\Delta_{A}-(\omega^{2}-
\omega_{P}^2)^{1/2})z} &  0\cr \gamma \cos\theta \sin\theta & 0 & 1-\gamma
\cos^{2}\theta} \right)\ .
\label{eq:pz}
\end{equation}
In the above equations, $\Delta = \Delta_{\phi} - \Delta_{A}$, while $\gamma = (1-e^{i\Delta z}$). $\Delta_{A}$ and $\Delta_{\phi}$ are defined in terms of the frequency, $\omega$, and the eigenvalues, $\mu_{\pm}^2$, as
\begin{eqnarray}
	\Delta_{A} = \sqrt{\omega^2-\mu_{+}^2}-\omega, \nonumber \\
	\Delta_{\phi} = \sqrt{\omega^2-\mu_{-}^2}-\omega\, .
\end{eqnarray}
We may approximate $\Delta$ as
\begin{equation}
	\Delta = \Delta_{\phi} - \Delta_{A} \approx \frac{1}{l} \sqrt{1 + \tan^{2}2\theta} \approx \frac{1}{l}.
\end{equation}
The approximation, applicable to our work here, is that $\omega \gg \omega_P,\ m_{\phi}$ and $g_{\phi}{\cal B}_T$. We also find that in our case $\tan 2\theta \ll 1$, as we elaborate below. For later reference, we give the leading term in the expansion of the phase in the 22 element of Eq. (\ref{eq:pz}), which we will denote $\delta$z, in terms of $\sin \theta$, of order 10$^{-5}$ in our application:
\begin{equation}
	i\delta z \equiv i(\omega+\Delta_{A}-(\omega^{2}-\omega_{P}^2)^{1/2})z \approx -i \frac{z}{l} \sin^2\theta ,
\label{eq:phase}
\end{equation}
where $\omega_P \ll \omega$ for $n_e \approx 10^{-8} \ \textrm{cm}^{-3}$, as commonly estimated \cite{Csaki}, and $\omega \sim$ 50 GHz for the CMB. The consequent expansion of the 22 element of $P(z)$ reads
\begin{equation}
 e^{-i \delta z} \approx 1 + i  \frac{z}{l} \sin^2\theta .
\label{eq:22}
\end{equation}

In the model that we consider, the medium has a uniform value of $\omega_{P}$, namely uniform electron density, and the cluster diameters are all the same, with value $z$, over which the direction and strength of the magnetic field is assumed to be uniform.  Each cluster has a random orientation with respect to some ``external" coordinate system, where the propagation direction is $z$.  Eq. (\ref{eq:rhoz}) may be taken as the propagation from the initially prescribed polarization correlation matrix $\rho(0)$ through a distance $z$, in a uniform field of strength ${\cal B}_T$ that is oriented along the ``parallel" axis. We take the magnetic field in the first cluster to be aligned at an angle $\beta_{1}$ to the fixed external coordinate system. Therefore we rotate the initial state by an angle $\beta_{1}$ with rotation matrix $R(\beta_{1})$ to align the ``parallel" axis along the magnetic field direction in the first cluster. After a distance $z$, a new zone (cluster) is entered, where the field has the same strength  but a random orientation with respect to the first. At this point, after the propagation through the first cluster, the polarization state has changed. The new state is still expressed in terms of the rotated coordinate system, of course. So we rotate back by angle $\beta_{1}$ to return to the frame of the external coordinate system. Now the state is rotated by an angle $\beta_{2}$ with rotation matrix $R(\beta_{2})$ to align the new ``parallel" axis along the magnetic field direction in the second cluster. The propagation through distance $z$ in the second cluster results in a new polarization density matrix, expressed now in terms of the new rotated coordinates. Rotate back to the original coordinate system, then forward to the new (third) cluster's magnetic field direction. This process continues until the propagation wave reaches ``Earth", after, say, $n$ clusters have been traversed. The propagation then amounts to the following matrix product:
\begin{eqnarray}
	\rho_{n}(z) & = & R^{-1}(\beta_{n})P(z)R(\beta_{n})R^{-1}(\beta_{n-1})P(z)R(\beta_{n-1})R^{-1}(\beta_{n-2})...R^{-1}(\beta_{1})P(z)R(\beta_{1}) \cr  
	&  & \times \; \rho(0)R^{-1}(\beta_{1})P^{-1}(z)R(\beta_{1})R^{-1}(\beta_{2})P^{-1}(z)R(\beta_{2})...R^{-1}(\beta_{n})P^{-1}(z)R(\beta_{n}),                    
\label{eq:rhon}
\end{eqnarray}
where the rotation matrix $R(\beta_{m})$ acts only on the two dimensional space transverse to the propagation direction, and it reads

\begin{equation}
	R(\beta_{m}) = \left(\matrix{\cos\beta_{m} & \sin\beta_{m} & 0 \cr
	-\sin\beta_{m} & \cos \beta_{m} & 0 \cr 0 & 0 & 1} \right)\ .
\label{eq:rot}
\end{equation}
Further simplifications occur, because $R(\beta_{m})R^{-1}(\beta_{m-1})$ = $R(\beta_{m}-\beta_{m-1})$, since $R^{-1}(\beta$)=$R(-\beta)$.  The reference axis for $\beta$ is arbitrary, since only the relative angles between the field directions in adjoining clusters are relevant. The difference between two random angles is obviously a random angle, so the structure of Eq. (\ref{eq:rhon}) amounts to the unitary transformation $\rho_{n}(z)$ = $U(z,n)\rho(0)U^{-1}(z,n)$, where $U(z,n)$ is the product of random rotations of $P(z)$. Next, since the parameter $\theta$ is very small ($\approx 10^{-5}$ in our case), we develop Eq. (\ref{eq:rhon}) to the leading order in $\theta$.

To expand the propagation matrix in powers of $\sin\theta$, write it as the sum of a (nearly) unit matrix and a matrix proportional to $\sin\theta$:
\begin{equation}
	P(z) \equiv \mathcal{I}(z) + \sin\theta \; \mathcal{P}(z,\theta).
\label{eq:defP}
\end{equation}
Suppressing the irrelevant over all phase in Eq. (\ref{eq:pz}),
% and the order tan$^2(2\theta)$ $\approx$ sin$^2(2\theta)$ correction to unity in the 2-2 element of $\mathcal{I}(z)$,
the matrices $\mathcal{I}(z)$ and $\mathcal{P}(z,\theta)$ are given by
\begin{equation}
	\mathcal{I}(z) =  \left(\matrix{1 & 0 & 0 \cr 0 & 1 &  0\cr 0 & 0 & e^{i\Delta z}} \right)\,
\label{eq:unit}
\end{equation}
and
\begin{equation}
	\mathcal{P}(z,\theta) = \left(\matrix{-\gamma \sin\theta  & 0 & \gamma \cos\theta  \cr 
	0 & i{z\over l}\sin\theta &  0 \cr 
	\gamma \cos\theta  & 0 & \gamma \sin\theta } \right)\ ,
\label{eq:calP}
\end{equation}
and $\gamma$ is defined as in Eq. (\ref{eq:pz}). Using the fact that $\mathcal{I}(z)$ and $R(\beta)$ commute and $\mathcal{I}(z)^{-1}$ and $R(\beta)^{-1}$ = $\mathcal{I}(-z)$ and $R(-\beta)$, respectively, one finds that to leading order in $\sin\theta \; \mathcal{P}(z,\theta)$,
\begin{eqnarray}
	\rho_{n}^{[1]}(Z) & = & \mathcal{I}(z)^{n}\rho(0)\mathcal{I}(-z)^n \cr
	& + & \sin\theta \sum_{j=1}^{n}\left[\mathcal{I}(z)^{n-j}R(-\beta_{j})\mathcal{P}(z,\theta)R(\beta_{j})\mathcal{I}(z)^{j-1}\rho(0)\mathcal{I}(-z)^{n}
 + h.c.\right],
\label{eq:leading}
\end{eqnarray}
where $Z=nz$ is the total distance travelled. The form of $\rho_{n}\approx \rho_{n}^{[1]}$, Eq. (\ref{eq:leading}), a leading zeroth order term and a sum of pairs of first order terms, each corresponding to non trivial propagation in only one of the $n$ cells, is what one expects to leading order. The above treatment of the model is perfectly general. 

We next assume that the electromagnetic wave is initially unpolarized. Furthermore we assume that the initial correlators,
\begin{equation}
<\phi(0) A_{||}^{*}(0)> \; = \; <\phi(0) A_{\bot}^{*}(0)> \; = \;
<\phi(0) \phi^{*}(0)> \; = \; 0\, .
\end{equation}
Hence the initial density matrix,
\begin{equation}
\rho(0) = \textrm{diag}(1,1,0).
\label{eq:initial_rho}
\end{equation}
In this case the order $\sin\theta$ terms, as defined in Eqs. (\ref{eq:defP}) and (\ref{eq:calP}), are actually of order $\sin^{2}\theta$ in the terms $\rho(z)_{11}$, $\rho(z)_{12}$, $\rho(z)_{21}$ and $\rho(z)_{22}$ that affect the polarization parameters at first order in $\sin\theta \; \mathcal{P}(z,\theta)$. The $ j^{th}$ term of the sum in the expression for $\rho_{n}^{[1]}$ reads, for the special case where $\rho(0) = (1,1,0)$,
\begin{equation}
\rho_{n}^{[1]}(z)_{j} =  \left(\matrix{-2\sin^2\theta \cos^2\beta_j(1-
\cos\Delta z) & -\sin^2\theta \sin2\beta_j (1-\cos\Delta z)& \gamma^*
e^{-i\Delta z(n-j)} \cos\beta_j \sin(2\theta)/2\cr  -\sin^2\theta \sin2\beta_j
(1-\cos\Delta z) & -2\sin^2\theta \sin^2\beta_j(1-\cos\Delta z)& \gamma^*
e^{-i\Delta z(n-j)}\sin\beta_j \sin(2\theta)/2\cr \gamma e^{i\Delta z(n-j)}
\cos\beta_j \sin(2\theta)/2 & \gamma e^{i\Delta z(n-j)}\sin\beta_j \sin(2\theta)/2 & 0} \right)\ ,
\label{eq:rho1}
\end{equation}
At second order in $\sin\theta \; \mathcal{P}(z,\theta)$, the mixing between A$_{||}$ and $\phi$, and A$_{\bot}$ and $\phi$ induces order $\sin^2\theta$ terms, so these must be included as well, and we deal with them next.  

There are $2n(2n-1)/2$ ways to choose two propagation factors $\mathcal{P}(z,\theta)$ from the $2n$ positions in the general expression Eq. (\ref{eq:rhon}). The general second order expression is an extension of the form shown in
Eq. (\ref{eq:leading}). Denoting it $\rho_{n}^{[2]}$, we find
\begin{eqnarray}
\rho_{n}^{[2]}(Z) & = &
\sin^2\theta \Big\{ \sum_{k=2}^{n}\sum_{j=1}^{k-1}\big[\mathcal{I}(z)^{n-k}R(-\beta_{k})\mathcal{P}(z,\theta)\mathcal{I}(z)^{k-j-1}R(\beta_{k}-\beta_{j})\mathcal{P}(z,\theta)R(\beta_{j})\mathcal{I}(z)^{j-1} \cr
&\times &\rho(0)\mathcal{I}(-z)^{n} + h.c. \big]\cr
& + &\sum_{k=1}^{n}\sum_{j=1}^{n}[\mathcal{I}(z)^{n-j}R(-\beta_{j})\mathcal{P}(z,\theta)R(\beta_{j})\mathcal{I}(z)^{j-1}\rho(0) \cr
&\times&\mathcal{I}(-z)^{k-1}R(-\beta_{k})\mathcal{P}(-z,\theta)R(\beta_{k})\mathcal{I}(-z)^{n-k}]\Big\}.
\label{eq:nextleading}
\end{eqnarray}
Using Eq. (\ref{eq:unit}), $\mathcal{I}(z)^m$ is simply diag(1,1,$e^{i\Delta mz})$, so $R(\beta)$ and $\mathcal{I}(z)$ commute. Thus their orders in Eqs. (\ref{eq:leading}) and (\ref{eq:nextleading}) are immaterial and can be
chosen for convenience.  Only the propagation matrix $\mathcal{P}(z,\theta)$  and, in principle, the initial conditions matrix, $\rho(0)$, do not generally commute with each other and with the  $R(\beta)$ and $\mathcal{I}(z)$ matrix.  In general their positions in the matrix product are fixed, though for special cases such as ours, which has the unpolarized initial condition with zero value for the $\phi$ field initially, $\rho(0)$ = diag(1,1,0), the matrices can be collapsed and simplified even further, as illustrated in Eq. (\ref{eq:rho1}) and below in Eq. (\ref{eq:nextleadingjk}).

When $\rho(0)$ = diag(1,1,0),  the $\rho(z)$ elements relevant to our study of spontaneous polarization  at leading order are of the form $\gamma^2 \sin^2\theta$ $\times$ products of cosines and sines of rotation angles. Corrections to the leading terms are themselves of order $\sin^2\theta$ $\approx$ $10^{-10}$ for the parameters we choose for our model.  Assembling all of the contributions of terms of second order in $\sin\theta \; {\cal P}(z,\theta)$, we find that the $j$-, $k$- th term of the sum in Eq. (\ref{eq:nextleading}), yields the following elements of the 2 $\times$ 2 submatrix relevant to polarization:
\begin{eqnarray}
\rho_{n}^{[2]}(z,j,k)_{11}  & =  &
-4 \sin^2\theta(1-\cos\Delta z)\cos[\Delta z(k-j)] \cos\beta_k \cos\beta_j
\nonumber \\
\rho_{n}^{[2]}(z,j,k)_{12} & =  &
-2 \sin^2\theta(1-\cos\Delta z)\cos[\Delta z(k-j)] (\sin(\beta_k+\beta_j)-
i\sin(\beta_k-\beta_j)\tan[\Delta z(k-j)])
\nonumber\\
\rho_{n}^{[2]}(z,j,k)_{21} & = &
-2 \sin^2\theta(1-\cos\Delta z)\cos[\Delta z(k-j)] (\sin(\beta_k+\beta_j)+
i\sin(\beta_k-\beta_j)\tan[\Delta z(k-j)])\nonumber\\
\rho_{n}^{[2]}(z,j,k)_{22} & = &
-4 \sin^2\theta(1-\cos\Delta z)\cos[\Delta z(k-j)] \sin\beta_k \sin\beta_j .
\label{eq:nextleadingjk}
\end{eqnarray}
It turns out that all of these contributions come from the terms in the first double sum in Eq. (\ref{eq:nextleading}), since the zero in the 33 entry of $\rho(0)$ prevents the order one elements 13, 23, 31 and 32 in $\rho_n^{[1]}(z)_j $ from communicating with those in $\rho_n^{[1]}(z)_k$ to form order one elements in the 11, 22, 12 and 21 product in the second double sum term in Eq. (\ref{eq:nextleading}).

At this point, combining each term with its hermitian conjugate, we have displayed $n(n-1)/2+n = n^2/2 + n/2$  
terms at order $\sin^2\theta$. It is this second order term in $\sin\theta \; {\cal P}(z,\theta)$ that dominates the CMB photon's random walk through the clusters. Inspection of Eq. (\ref{eq:nextleadingjk}) reveals that the polarization parameters $Q$, $U$ and $V$ are, on average, all the same order of magnitude. Our analysis leads to the conclusion that the number of contributing ``steps" in the random walk represented by the propagation process is proportional to $n^2$, so the effective displacement will be proportional to $n$, with the spontaneous polarization at a step being of order $10^{-10}$. The dependence of the intensity on the number of domains has been discussed earlier \cite{Csaki,sroy,MirizziCMB,MirizziTEV}. However the linear dependence of polarization observables $Q$, $U$ and $V$ on $n$ is a new result. For the transport through a thousand cells, or clusters, we expect a typical net displacement (polarization) of order $10^{-7}$, given our parameter assumptions.  

We can compute reduced Stokes parameters using the correlation functions of $A_{||}(z)$ and $\phi(z)$,
\begin{eqnarray}
	I & = & <A_{||}(z) A_{||}^{*}(z)> + <A_{\bot}(z) A_{\bot}^{*}(z)> = \rho_{11} + \rho_{22}, \\
	Q & = & <A_{||}(z) A_{||}^{*}(z)> - <A_{\bot}(z) A_{\bot}^{*}(z)> = \rho_{11}-\rho_{22}, \\
	U & = & <A_{\bot}(z) A_{||}^{*}(z)> + <A_{||}(z) A_{\bot}^{*}(z)>  = \rho_{12} + \rho_{21}, \\
	V & = & i(- <A_{\bot}(z) A_{||}^*(z)> + <A_{||}(z) A_{\bot}^{*}(z)> ) = i(- \rho_{12} + \rho_{21}).
\end{eqnarray}
The linear polarization angle is given by,
\begin{equation}
	\tan 2\psi = \frac{U}{Q},
\end{equation}
and the degree of polarization is given by,
\begin{equation}
	p = \frac{\sqrt{Q^2 + U^2 + V^2}}{I}.
\end{equation}
With this background analysis of the model, we turn to applications to CMB temperature and polarization.

%%%%%%%%%%%%%%%%%%%%%%%%%%%%%%%%%%%%%%%%%%%%%%%%%%%%%%%%%%%%%%%%%

%%%%%%%%%%%%%%%%%%%%%%%%%%%%%%%%%%%%%%%%%%%
\section{CMB Temperature and Polarization}
%%%%%%%%%%%%%%%%%%%%%%%%%%%%%%%%%%%%%%%%%%%

In this section we consider the effect of pseudoscalar mixing on the propagation of CMB photons. The mixing affects both the CMB temperature and polarization. We can impose a limit on the pseudoscalar-photon coupling by requiring that the CMB is distorted to less than one part in $10^{5}$, which is the amplitude of CMB temperature fluctuations. Furthermore we also demand that the mixing does not generate a CMB polarization larger than $10^{-6}$. Finally since the WMAP has not observed $B$ modes at the level of $10^{-6}$, we require that these are at least an order of magnitude smaller. The limit can be imposed only if we assume a value for the background magnetic field and plasma density. The magnetic field, in particular, is very poorly known. Furthermore the results can change significantly if the magnetic field and plasma density changes along the path as well as if there is a background flux of pseudoscalars. We set the incident pseudoscalar flux to zero. Furthermore we assume a uniform background field and plasma density within a particular domain. The result is also found to depend significantly on the domain size that we assume and hence on the details of how we model in the intergalactic medium. Due to the uncertainty in the magnitude of the magnetic field it is best to impose a limit on the product $g_\phi {\cal B}_T$. This is possible since the magnetic field always occurs in the form of this product. The model dependence implies that the limit also depends on some parameters of the model, such as the domain size and/or the total number of domains. We shall address this issue later in this section.

The most stringent limits on the coupling $g_\phi$ are obtained from astrophysics \cite{PDG06}. The current astrophysical limit is $g_\phi<6\times 10^{-11}$ GeV$^{-1}$. The typical upper limit value of the intergalactic magnetic field is about $10^{-9}$ G  \cite{Kronberg} and the electron number density $n_e\approx 10^{-8}$ cm$^{-3}$ \cite{Csaki}. We point out that the supercluster magnetic
field may be significantly larger, of order $10^{-7}$ G \cite{Vallee}.
In broad outline, we model the intergalactic medium with several magnetic domains, each of size $z$ of order a few Mpc. The total propagation distance is of the order of a few Gpc. We shall assume that the magnetic field in different domains is uncorrelated.

Choosing specific values, we calculate the effect on CMB temperature and polarization observables as CMB radiation propagates through $n=1000$ magnetic domains, each of size $z=1$ Mpc. The relevant physical quantities are:
\begin{enumerate}
	\item Magnetic field, ${\cal B}_T = 10^{-9}$ G,
	\item Electron number density, $n_{e} = 10^{-8}\ \textrm{cm}^{-3}$,
	\item Frequency of radiation (CMB), $\nu = 50$ GHz, hence the oscillation length parameter, $l = 200$ pc.
\end{enumerate}
We set the coupling $g_\phi=6\times 10^{-11}$ GeV$^{-1}$. We numerically compute the Stokes parameters after propagating through $n$ clusters by using the equations given in Appendix A. After propagating through each cluster we rotate the electromagnetic wave vector in order to account for the change in the direction of the transverse magnetic field from one cluster to the next. For this purpose we need to transform the correlators in a fixed reference frame, at the begining of each cluster, to the reference frame with $x$-axis parallel to the transverse magnetic field in the cluster. The relevant formulas for the transformation of the correlators are given explicitly in Appendix B. We point out that the initial density matrix is given by Eq. (\ref{eq:initial_rho}).

\begin{figure}
\includegraphics[width=3.5in,height=3.5in,angle=0]{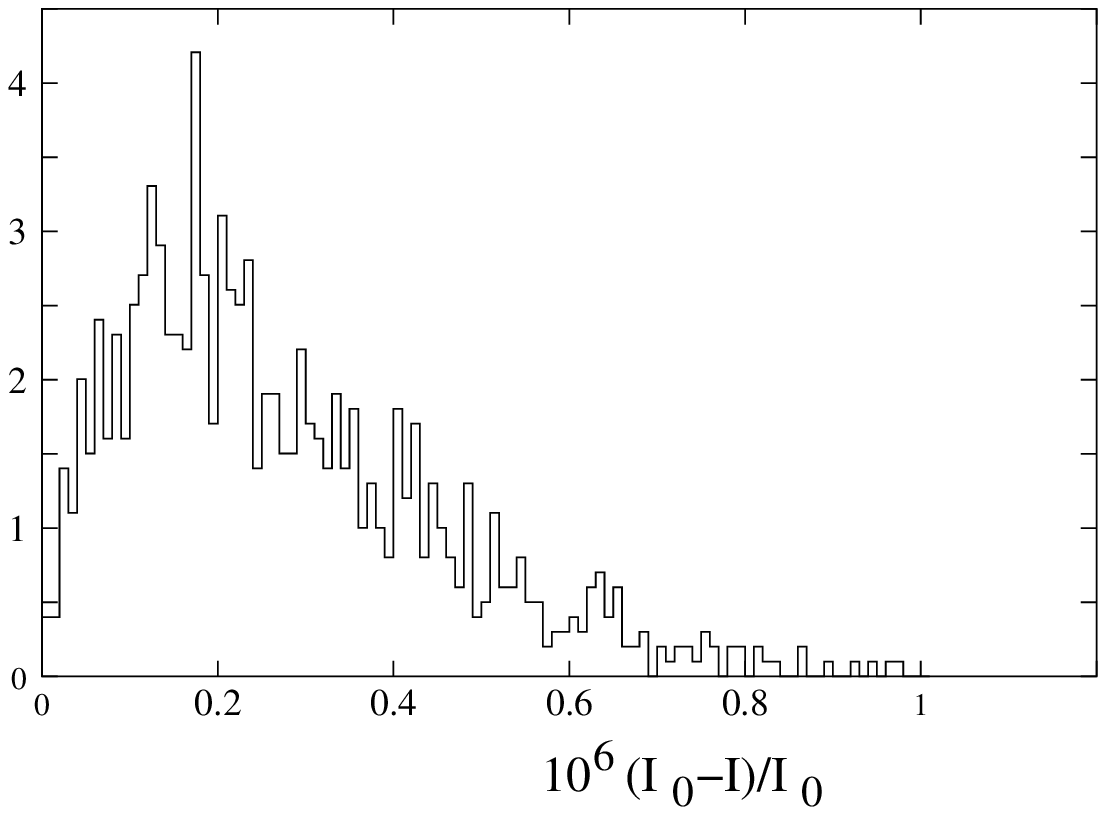}
\includegraphics[width=3.5in,height=3.5in,angle=0]{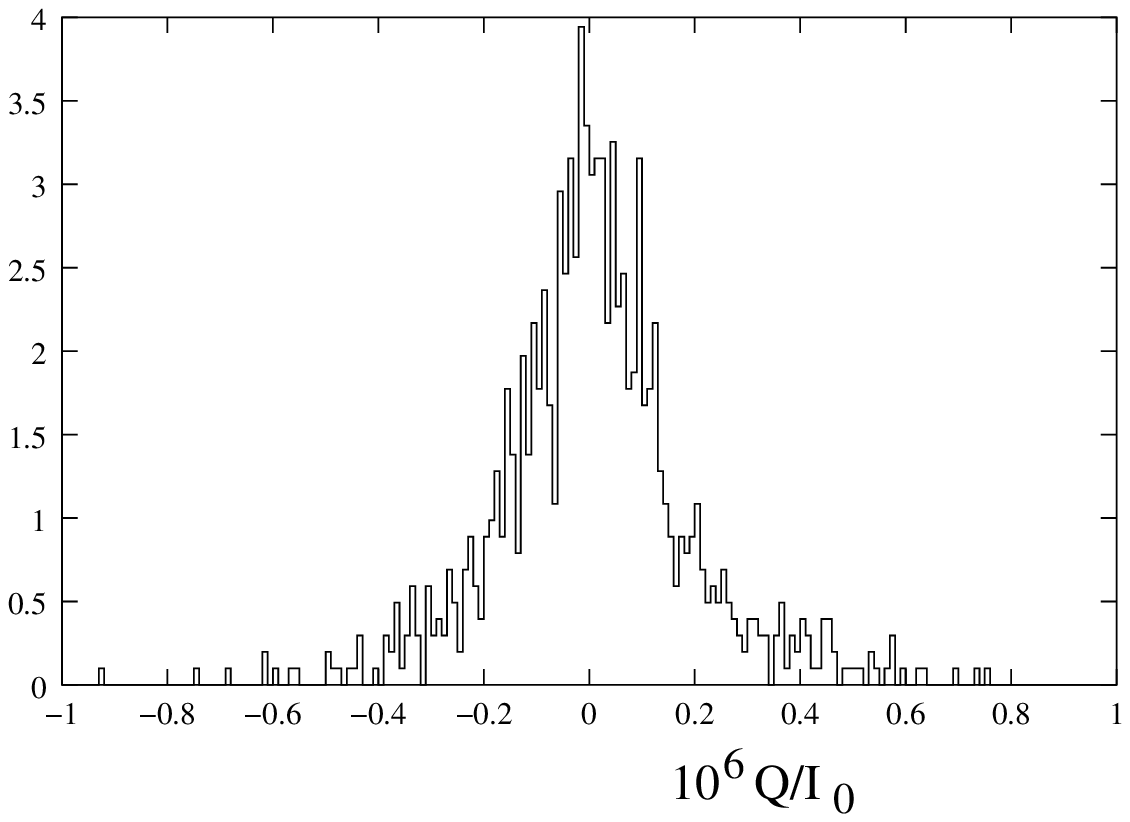}
\includegraphics[width=3.5in,height=3.5in,angle=0]{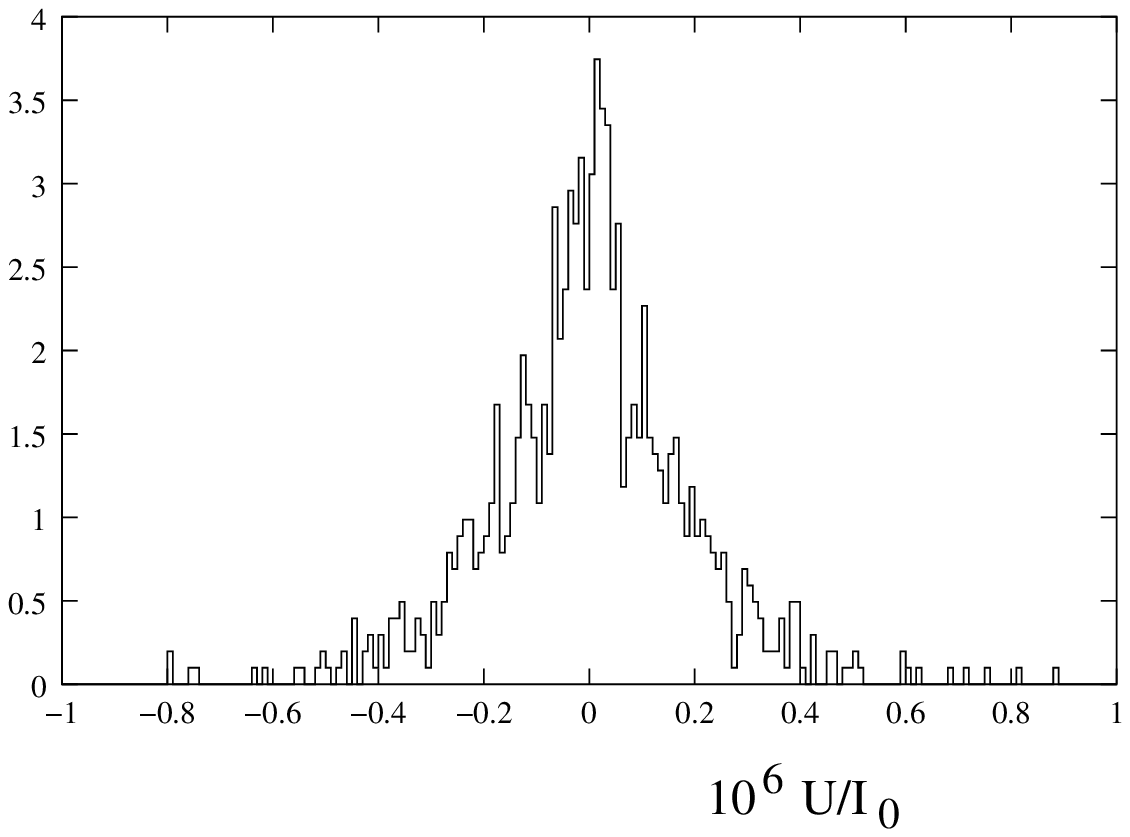}
\includegraphics[width=3.5in,height=3.5in,angle=0]{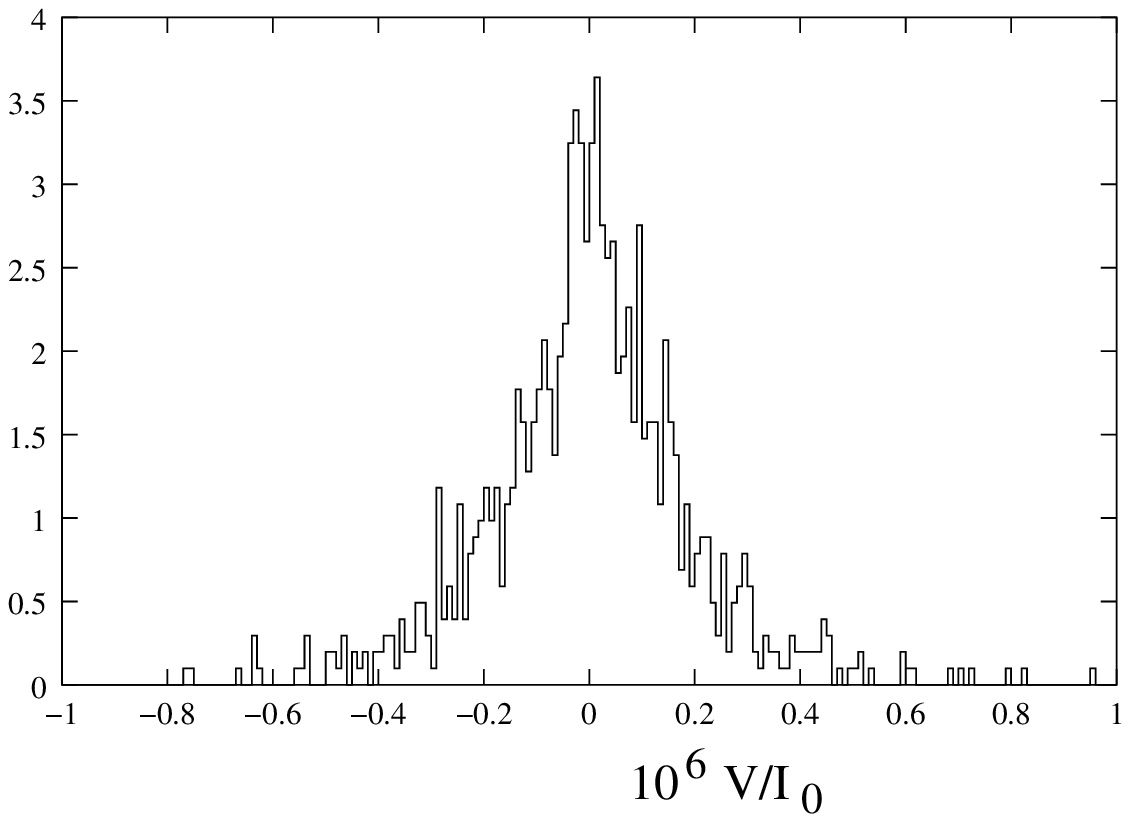}
\caption{The distribution, $f(x) \; (\times 10^{-6})$, of the normalized Stokes $I$, $Q$, $U$ and $V$ parameters for the CMB. For $I$ we only show the difference $I_0-I$, where $I_0$ is the initial intensity. All of the parameters have been normalized by dividing by $I_0$. Each histogram contains the results of 1000 simulations.}
\label{fig:Stokes}
\end{figure}

In Fig. \ref{fig:Stokes} we show histograms of 1000 simulations of $I_0-I$, $Q$, $U$ and $V$ Stokes parameters after propagating over 1000 clusters. All the parameters are normalized by the initial intensity $I_0$. Here we have assumed that initially the wave is unpolarized. If initially the wave has a non-zero value of the Stokes $Q$ parameter, our results remain essentially unchanged with the mean value of the $Q$ parameter shifted by its initial value. The effect on the $U$ and $V$ polarization is negligible.

In CMB studies one deals with the coordinate independent $E$ and $B$ modes. In the present case we work directly with the coordinate dependent $Q$ and $U$ Stokes parameters. This is fine as long as we are working in a particular direction in the sky. We align the coordinate system such that in the absence of pseudoscalar-photon mixing, only the $Q$ polarization is non-zero. Then the $U$ polarization generated by the mixing in the chosen frame provides an estimate of the $B$ modes. Hence the polarization generated by the pseudoscalar-photon mixing must be less than the current limit on the $B$ modes.

In Fig. \ref{fig:Stokes} the intensity is found to deviate from its inital values by less than 1 part in $10^{6}$ due to mixing for the chosen parameters. Hence the effect of mixing on temperature anisotropy is negligible. Also the Stokes parameter $Q$ remains very close to its initial value of zero. We see that the predictions are not in conflict with the current observations. We predict relatively large fluctuations in the $U$ parameter and the circular polarization $V$. The sample mean values of $U$ and $V$ are found to be approximately zero, typically of the order of $10^{-9}$. The corresponding standard deviations are found to be $2.0\times 10^{-7}$ for both of the parameters. Hence we find that the choice of parameters already predict the fluctuations in the $U$ polarization close to its limiting value. We, therefore, impose a limit on the product $g_\phi {\cal B}_T< 0.2$ Mpc$^{-1}$ due to CMB polarization.

\begin{figure}
\centering
\includegraphics[]{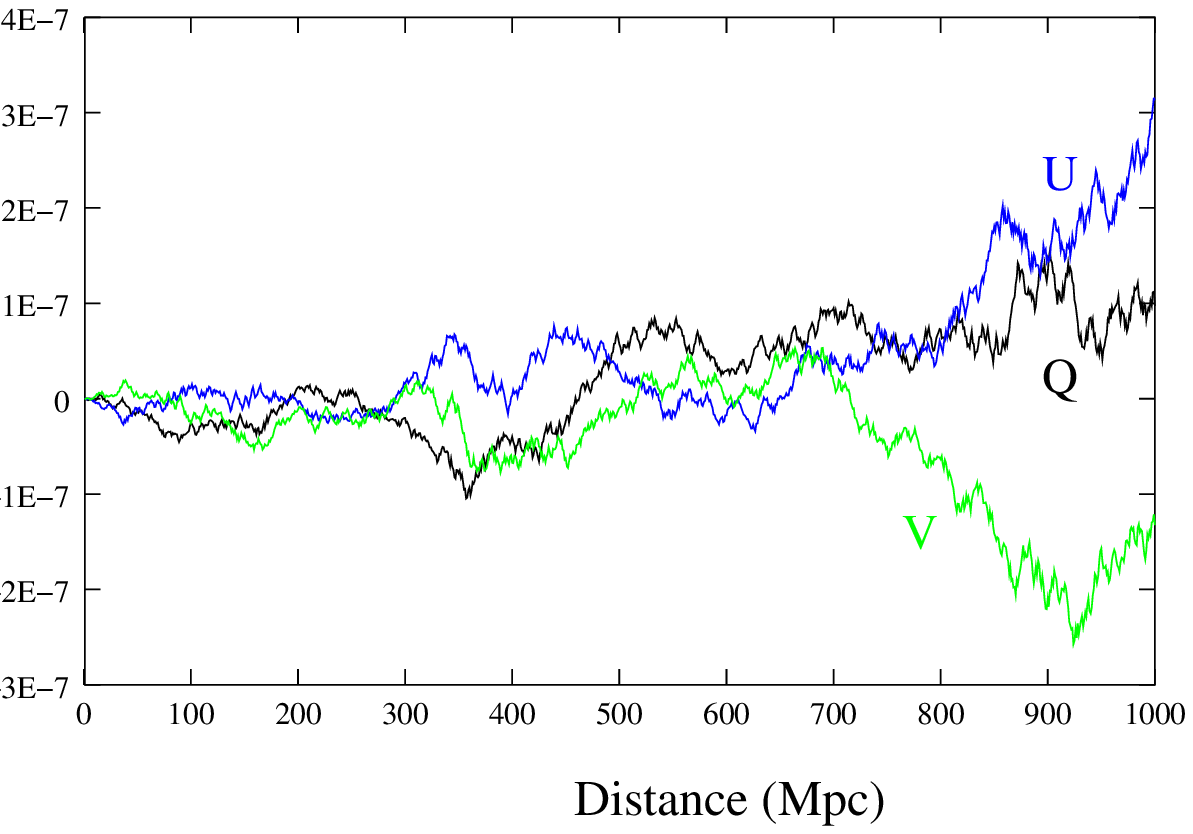}
\caption{The distance dependence of the normalized $Q$, $U$ and $V$ Stokes parameters for a particular random realization.}
\label{fig:xdep}
\end{figure}

In Fig. \ref{fig:xdep} we show the fluctuations in $Q$, $U$ and $V$ as a function of distance for a particular realization of the model.  The plot for this individual case shows similar magnitudes induced  for all the three Stokes parameters.  This is true for the entire distribution of realizations as well, as shown in Fig. \ref{fig:Stokes}.

\begin{figure}
\centering
\includegraphics[width=5in,height=3.5in,angle=0]{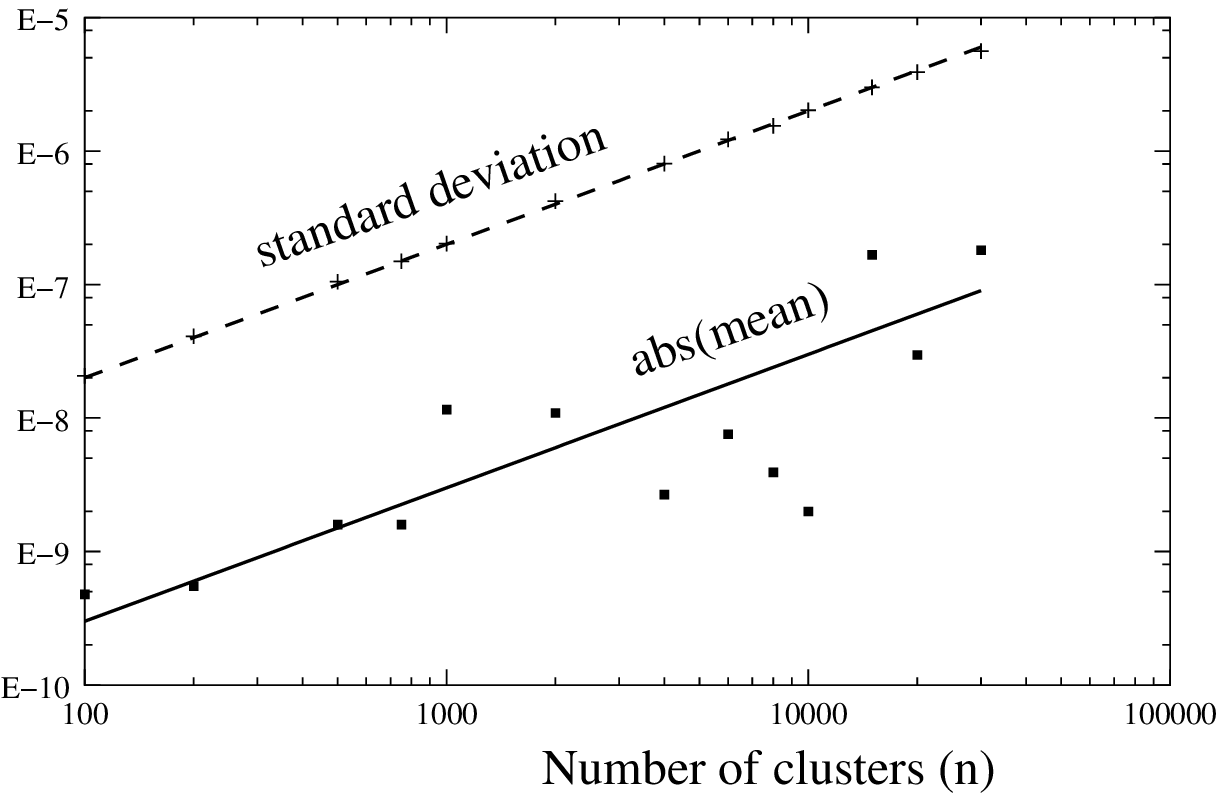}
\caption{The mean and standard deviation of $U/I_0$ as a function of the number of magnetic clusters $n$. For each $n$ we use a total of 1000 simulations. Both the mean (lower curve) and standard deviation (upper curve) show linear dependence on $n$.}
\label{fig:stats}
\end{figure}

With results of the full simulation in hand, we can check how the results depend on the number of clusters. In Fig. \ref{fig:stats} we plot the dependence of mean (absolute value) and standard deviation of $U$ on the number of clusters. The cluster size is kept fixed in this calculation. As we anticipate from our discussion in Sec. 2, we find that both the mean and standard deviation increase linearly with the number of clusters. The standard deviation shows very little digression from the linear plot, whereas the mean shows relatively large fluctuations. We point out that this linear behavior is expected as long as the Stokes parameters are much smaller compared to unity. If the number of clusters or the magnetic field in each cluster becomes too large then the linear dependence will break down.
 
\begin{figure}
\centering
\includegraphics[width=5in,height=3.5in,angle=0]{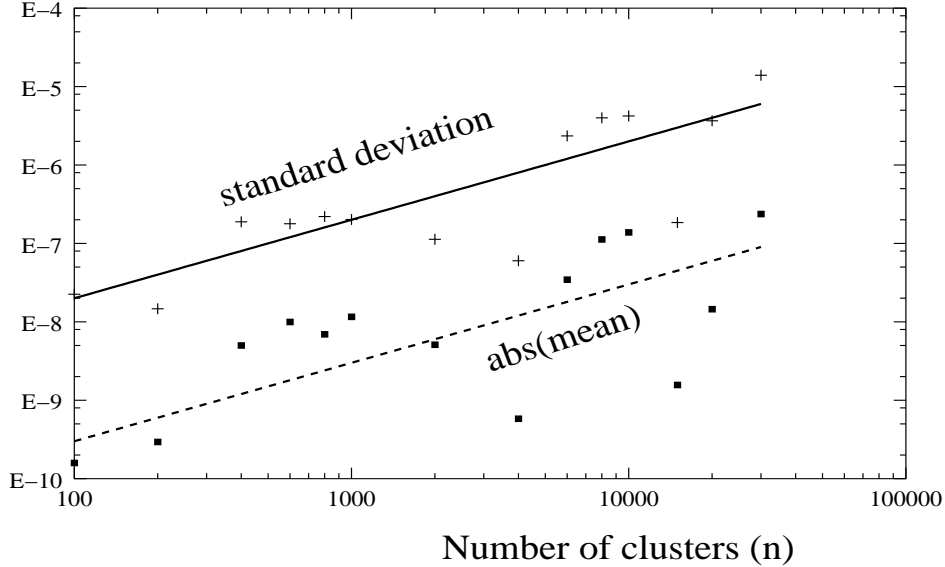}
\caption{The mean and standard deviation of $U/I_0$ as a function of the number of magnetic clusters $n$. Here the total distance of propagation is kept fixed. For each $n$ we use a total of 1000 simulations. Both the mean (lower curve) and standard deviation (upper curve) show roughly linear dependence on $n$.}
\label{fig:statsFD}
\end{figure}

We next determine how our predictions depend on the assumed model of intergalactic space. In Fig. \ref{fig:statsFD} we show how the mean and standard deviation of the $U$ parameter change as we reduce the size of each cluster and proportionately increase the total number of clusters such that the total distance of propagation remains unchanged. We find that the $U$ Stokes parameter increases as the cluster size is reduced. This is easily understood. The oscillation length $l$ is much smaller than the size of each cluster. Hence, statistically, the contribution per cluster remains almost unchanged as we decrease the size $z$ of the cluster as long as $z \gg l$. This is clear from the analytical result, Eq. (\ref{eq:nextleadingjk}), obtained in Sec. 2. The result depends on $z$ only through the functions such as $\cos\Delta z$, $\tan[\Delta z (k-j)]$. Since the arguments of the trigonometric functions are much larger than unity, adding a large number of such terms yields a negligible or fluctuating dependence on $z$. However the total number of clusters, $n$, keeps increasing, leading to a linear increase with $n$ in the total contribution, as explained in Sec. 2. Due to this model dependence we may express the limit more reliably as
\begin{equation}
g_\phi {\cal B}_T < \sqrt{1000\over n} \times 0.2\ {\rm Mpc}^{-1}.
\label{eq:limit}
\end{equation}
We expect this to be valid as long as the cluster size, $z$, is much larger than $1/\Delta$. Although we have obtained this assuming uniform cluster size, and magnetic field strength equal in all clusters, the result is approximately valid even if the magnetic field and cluster size fluctuates from cluster to cluster. In this case we interpret the magnetic field strength in Eq. (\ref{eq:limit}) as the mean magnetic field over all the clusters. We have explicitly verified this numerically by allowing fluctuations in the magnetic field and cluster size with standard deviation of the order of these parameters. However the result may deviate significantly if these parameters show much larger fluctuations.

\begin{figure}
\centering
\includegraphics[width=5in,height=3.5in,angle=0]{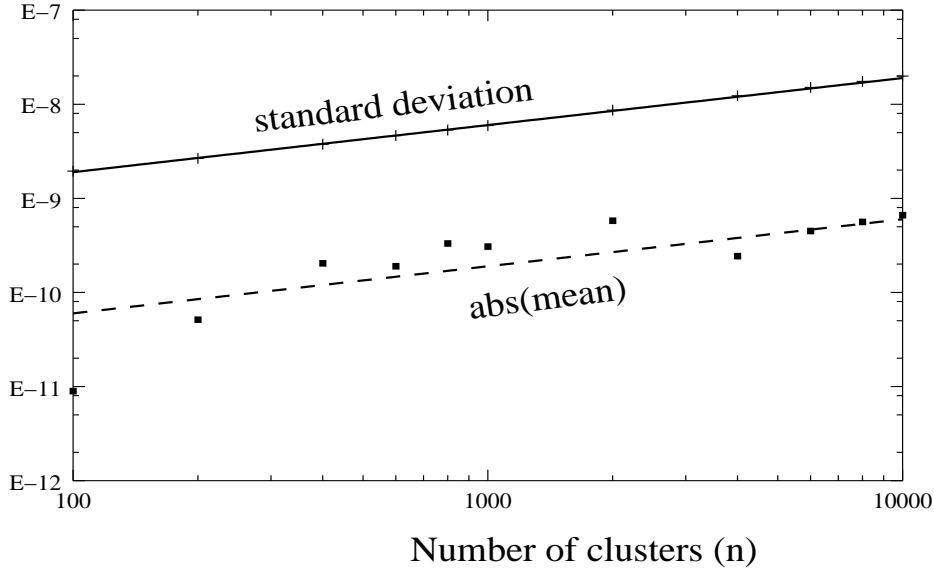}
\caption{The mean and standard deviation of $U/I_0$ as a function of the number of magnetic clusters $n$. Here all the correlations of $\phi$ with fields other than $\phi$ have been artificially set to zero at the beginning of each cluster. For each $n$ we use a total of 1000 simulations. Both the mean (lower curve) and standard deviation (upper curve) show linear dependence on $\sqrt{n}$.
}
\label{fig:test}
\end{figure}

What part of the density matrix is driving the linear growth with the number of clusters $n$? From our discussion in Sec. 2, we see that the linear growth comes from the second order terms which involve the correlators of $\phi$ with $A_{||}$ and $A_\bot$. We can verify this numerically by setting all such correlators to zero at each step. In Fig. \ref{fig:test} we show the result of this numerical experiment. The consequence is a growth in the Stokes parameter $U$ proportional to $\sqrt n$. This confirms our expectation in Sec. 2 that the linear growth is driven by the correlators $<\phi(z) A^*_{||}(z)>$ and $<\phi(z) A^*_\bot(z)>$.

The values of the $U$ and $V$ Stokes parameters are found to be relatively large. Such large values of the $U$ and $V$ parameters are somewhat surprising and it is useful to get an independent estimate to verify the numerical results. For the parameters chosen the mixing angle $|\theta|\approx g_\phi {\cal B}_T l/2 \approx 10^{-5} $. Furthermore the oscillation length $l\approx 10^{-4}$ Mpc $\ll z$. Hence we expect that in one domain the $U$ polarization would be of order $\theta^2\approx 10^{-10}$. As we have seen above the effect is linearly proportional to the number of clusters. Hence for 1000 clusters we multiply this number by 1000 to get $10^{-7}$, in agreement with the value found by direct
numerical computation.

%%%%%%%%%%%%%%%%%%%%%%%%%%%%%%%%%%%%%%%%%%%%%%%%%%%%%%%%%%%%%%%%%

%%%%%%%%%%%%%%%%%%%%%%%%%%%%%%%%%
\section{Summary and discussion}
%%%%%%%%%%%%%%%%%%%%%%%%%%%%%%%%%

Proposed as an extension to the standard model in order to solve the strong-CP problem in quantum chromodynamics (QCD), axions today still remain undetected. Direct and indirect (such as ours) methods of detection are being used to search for axion-like particles.
%If observed, they could possibly explain the origins of dark matter.
Future missions such as Planck will improve the sensitivity to the polarization of the CMB, which is a crucial next step in pinning down the origin of anisotropies in the CMB. A detection of circular polarization would certainly strengthen the case for existence of ultra-light pseudoscalars (such as axions).

In this paper we have analyzed the mixing of pseudoscalars and photons in order to explain the polarization of the cosmic microwave background radiation. We have treated the intergalactic region as a collection of magnetic domains (each of size 1 Mpc), and have shown, both theoretically and numerically, that the polarization of radiation increases linearly with the number of clusters (domains). Within the experimentally allowed range of $g_\phi {\cal B}_T$ values, our model predicts a $B$ mode polarization and circular polarization up to order $10^{-7}$.  

The model that we have used assumes that magnetic fields in adjoining clusters are completely uncorrelated. Although this is the most reasonable assumption, it would be interesting to see how things change if the fields are instead correlated. We might, for example, be able to explain the reported observation of large scale alignment of the optical polarization of quasars \cite{huts}. We intend to address this issue in a future publication.

%%%%%%%%%%%%%%%%%%%%%%%%%%%%%%%%%%%%%%%%%%%%%%%%%%%%%%%%%%%%%%%%%

%%%%%%%%%%%%%%%%%%%%%%
\section*{Appendix A}
%%%%%%%%%%%%%%%%%%%%%%

Spelled out in components, the propagation equation of the density matrix $\rho(z)$ amounts to six independent equations:

\begin{eqnarray}
<A_{||}(z) A_{||}^{*}(z)> & = & \frac{1}{2} <A_{||}(0) A_{||}^{*}(0)>
 \Bigl[ 1 + \textrm{cos}^{2} \; 2\theta + \textrm{sin}^{2} \; 2\theta \;
       \textrm{cos}[z(\Delta_{\phi} - \Delta_{A})] \Bigr]  \nonumber \\
        & + & \frac{1}{2} <\phi(0) \phi^{*}(0)>
\Bigl[ \textrm{sin}^{2} \; 2\theta - \textrm{sin}^{2} \; 2\theta \; \textrm{cos}[z(\Delta_{\phi} - \Delta_{A})]   \Bigr]  \nonumber \\
        & + & \Biggl\{ \frac{1}{2} <\phi(0) A_{||}^{*}(0)>
              \Bigl[ \textrm{sin} \; 2\theta \; \textrm{cos} \; 2\theta - \textrm{sin} \; 2\theta \; \textrm{cos} \; 2\theta                                                 \; \textrm{cos}[z(\Delta_{\phi} - \Delta_{A})] \nonumber \\
        & - &        \textrm{i} \; \textrm{sin} \; 2\theta \; \textrm{sin}[z(\Delta_{\phi} - \Delta_{A})]
     \Bigr] + \textrm{c.c.}
              \Biggr\}  \\                                              
        <A_{\bot}(z) A_{\bot}^{*}(z)> & = & <A_{\bot}(0) A_{\bot}^{*}(0)>  \\
        <A_{\bot}(z) A_{||}^{*}(z)>   & = & <A_{\bot}(0) A_{||}^{*}(0)>        
        \Bigl[ \textrm{cos}^{2} \; \theta \; \textrm{e}^{\textrm{\scriptsize i}\;Fz} + \textrm{sin}^{2} \; \theta \;                                                   \textrm{e}^{\textrm{\scriptsize i}\;Gz}
        \Bigr]  \nonumber \\
        & + & <A_{\bot}(0) \phi^{*}(0)>        
        \Bigl[ \textrm{sin} \; \theta \; \textrm{cos} \; \theta
    \left( \textrm{e}^{\textrm{\scriptsize i}\;Fz} - \textrm{e}^{\textrm{\scriptsize i}\;Gz}
      \right)\Bigr]  \\
<\phi(z) \phi^{*}(z)>         & = & \frac{1}{2} <A_{||}(0) A_{||}^{*}(0)>
      \Bigl[ \textrm{sin}^{2} \; 2\theta - \textrm{sin}^{2} \; 2\theta \; \textrm{cos}[z(\Delta_{\phi} - \Delta_{A})]
              \Bigr]  \nonumber \\
        & + & \frac{1}{2} <\phi(0) \phi^{*}(0)>
      \Bigl[ 1 + \textrm{cos}^{2} \; 2\theta + \textrm{sin}^{2} \; 2\theta \;
       \textrm{cos}[z(\Delta_{\phi} - \Delta_{A})]                                    \Bigr]  \nonumber \\ & + & \Biggl\{ \frac{1}{2} <\phi(0) A_{||}^{*}(0)>
   \Bigl[ - \textrm{sin} \; 2\theta \; \textrm{cos} \; 2\theta + \textrm{sin} \; 2\theta \; \textrm{cos} \; 2\theta                                                 \; \textrm{cos}[z(\Delta_{\phi} - \Delta_{A})] \nonumber \\
& + &   \textrm{i} \; \textrm{sin} \; 2\theta \; \textrm{sin}[z(\Delta_{\phi} - \Delta_{A})]       \Bigr] + \textrm{c.c.} \Biggr\}  \\
<\phi(z) A_{||}^{*}(z)>       & = & \frac{1}{2} <A_{||}(0) A_{||}^{*}(0)>
\Bigl[ \textrm{sin} \; 2\theta \; \textrm{cos} \; 2\theta - \textrm{sin} \; 2\theta \; \textrm{cos} \; 2\theta                                                 \; \textrm{cos}[z(\Delta_{\phi} - \Delta_{A})] \nonumber \\
& - &        \textrm{i} \; \textrm{sin} \; 2\theta \; \textrm{sin}[z(\Delta_{\phi} - \Delta_{A})]  \Bigr]  \nonumber \\
& + & \frac{1}{2} <\phi(0) \phi^{*}(0)>
\Bigl[ - \textrm{sin} \; 2\theta \; \textrm{cos} \; 2\theta + \textrm{sin} \; 2\theta \; \textrm{cos} \; 2\theta                                                 \; \textrm{cos}[z(\Delta_{\phi} - \Delta_{A})] \nonumber \\
        & + &          \textrm{i} \; \textrm{sin} \; 2\theta \; \textrm{sin}[z(\Delta_{\phi} - \Delta_{A})]      \Bigr]  \nonumber \\         & + & \frac{1}{2}
        <\phi(0) A_{||}^{*}(0)>        \Bigl[ \textrm{sin}^{2} \; 2\theta + \left( 1 + \textrm{cos}^{2} \; 2\theta \right) \textrm{cos}[z(\Delta_{\phi}                                               - \Delta_{A})]  \nonumber \\
& + &        \textrm{2i} \; \textrm{cos} \; 2\theta \; \textrm{sin}[z(\Delta_{\phi} - \Delta_{A})]         \Bigr]  \nonumber \\
& + & \frac{1}{2} <A_{||}(0) \phi^{*}(0)>
\Bigl[ \textrm{sin}^{2} \; 2\theta - \textrm{sin}^{2} \; 2\theta \; \textrm{cos}[z(\Delta_{\phi} - \Delta_{A})]\Bigr] \\
<\phi(z) A_{\bot}^{*}(z)>            & = & <A_{||}(0) A_{\bot}^{*}(0)>
\Bigl[ \textrm{sin} \; \theta \; \textrm{cos} \; \theta
\left( \textrm{e}^{- \textrm{ {\scriptsize i}}\;Fz} - \textrm{e}^{- \textrm{ {\scriptsize i}}\;Gz}
 \right)        \Bigr]  \nonumber \\
& + & <\phi(0) A_{\bot}^{*}(0)>
\Bigl[ \textrm{sin}^{2} \; \theta \; \textrm{e}^{- \textrm{ {\scriptsize i}}\;Fz} + \textrm{cos}^{2} \; \theta \;                                              \textrm{e}^{- \textrm{ {\scriptsize i}}\;Gz} \Bigr],                
\end{eqnarray}
where
\begin{eqnarray}
        F  &\approx & \frac{1}{2\omega} (\mu_{+}^{2} - \omega_{P}^{2})\\
        G  &\approx & \frac{1}{2\omega} (\mu_{-}^{2} - \omega_{P}^{2})
\end{eqnarray}
and the remaining symbols are defined in Sec. 1 and 2 in the text.

%%%%%%%%%%%%%%%%%%%%%%
\section*{Appendix B}
%%%%%%%%%%%%%%%%%%%%%%

Let $A_1(n,0)$ and $A_2(n,0)$ be the two components of the electromagnetic field in a fixed reference frame at the begining of the $n^{th}$ cluster. For each cluster we define a local reference frame with $x$-axis aligned parallel to the transverse component of the background magnetic field. In the local frame corresponding to the  $n^{th}$ cluster we denote the two electromagnetic field components as $A_{||}(n,0)$ and $A_{\bot}(n,0)$. We assume that initially, i.e. at $n=0$, $z=0$, only the correlators $<A_{1}(0,0) A_{1}^{*}(0,0)>$ and $<A_{2}(0,0) A_{2}^{*}(0,0)>$ are nonzero. After propagation through the first cluster all the six correlation functions are likely to be non-zero. After transforming into the local frame, the initial correlators for the $n^{th}$ cluster are,
\begin{eqnarray}
<A_{||}(n,0) A_{||}^{*}(n,0)> & = & \textrm{cos}^{2} \; \alpha
<A_{1}(n,0) A_{1}^{*}(n,0)> + \; \textrm{sin}^{2} \; \alpha
<A_{2}(n,0) A_{2}^{*}(n,0)>  \nonumber \\
& + & \textrm{sin} \; \alpha \; \textrm{cos} \; \alpha \;
(<A_{2}(n,0) A_{1}^{*}(0)> + <A_{1}(n,0) A_{2}^{*}(0)>) ,  \\
<A_{\bot}(n,0) A_{\bot}^{*}(n,0)> & = & \textrm{sin}^{2} \; \alpha
<A_{1}(n,0) A_{1}^{*}(n,0) > + \; \textrm{cos}^{2} \; \alpha
<A_{2}(n,0) A_{2}^{*}(n,0)>  \nonumber \\
& - & \textrm{sin} \; \alpha \; \textrm{cos} \; \alpha \;
(<A_{2}(n,0) A_{1}^{*}(n,0)> + <A_{1}(n,0) A_{2}^{*}(n,0)>) ,                                         \\
<A_{\bot}(n,0) A_{||}^{*}(n,0)>   & = & \textrm{cos}^{2} \; \alpha
<A_{2}(n,0) A_{1}^{*}(n,0)> - \; \textrm{sin}^{2} \; \alpha
<A_{1}(n,0) A_{2}^{*}(n,0)>  \nonumber \\
& + & \textrm{sin} \; \alpha \; \textrm{cos} \; \alpha \;
(<A_{2}(n,0) A_{2}^{*}(n,0)> - <A_{1}(n,0) A_{1}^{*}(n,0)>)  ,                                       \\
<\phi(n,0) A_{||}^{*}(n,0)>  & = & \textrm{cos} \; \alpha
<\phi(n,0) A_{1}^{*}(n,0)> + \; \textrm{sin} \; \alpha
<\phi(n,0) A_{2}^{*}(n,0)> , \\
        <\phi(n,0) A_{\bot}^{*}(n,0)>   & = & - \; \textrm{sin} \; \alpha
<\phi(n,0) A_{1}^{*}(n,0)> + \; \textrm{cos} \; \alpha
<\phi(n,0) A_{2}^{*}(n,0)> ,
\end{eqnarray}
and
$<\phi(n,0) \phi^{*}(n,0)>$.

%%%%%%%%%%%%%%%%%%%%%%%%%%%%%%%%%%%%%%%%%%%%%%%%%%%%%%%%%%%%%%%%%%%%%%%%%%%%%%

\end{document}